\documentstyle[aps,prl,twocolumn]{revtex}
 \include{psfig}

\begin{document}
\draft
\preprint{ }
\title{Energy  and entropy of metastable states
in  glassy systems
}
\author{J. Christian Sch{\"o}n}
\address{Institut f{\"u}r Anorganische Chemie,
Universit{\"a}t Bonn\\
 Gerhard-Domagk-Str. 1, D-53121 Bonn 
}
\author{Paolo Sibani}
\address{Dept. of Physics,
Odense Universitet\\
 Campusvej 55, DK-5230 Odense M 
} 
\date{\today}
\maketitle
\begin{abstract}
We  investigate the multi-valley   energy
landscape  of a 3-D on-lattice  network model
  for covalent glasses,
  numerically determining 
the shape of the valleys,  the  local density of states,
 the density of minima
and the local connectivity. We present some of these quantities  in a 
graphical birds-eye view  of the landscape, and discuss 
their  implications  for the relaxation dynamics
and cooling behavior of glasses.
 The strong similarities between the 
landscape of this model and those of other
complex systems  point to the possibility   
of a common low-temperature dynamical  description.
\end{abstract}

 \narrowtext
 
  {\em \bf Introduction.}  
A  number  of problems ranging from  solid
state physics to combinatorial optimization\cite{Fischer91a},
 through chemical 
physics\cite{Berry93a,Ball96a,Becker97a,Wales98a},
and  biology\cite{Dobson98a,Landscape_paradigms}  can be
formulated in terms of Marcov processes with states in the 
systems' energy landscapes. 
The (effective)  connectivity of  the landscape specifies
 the allowed hopping transitions
while  the energies determine their magnitude.
Specifically  for the  multi-valley landscapes characterizing
complex systems,  local thermal equilibrium  establishes itself    
within `ergodic components'\cite{Palmer82a}, 
and an  investigation of  the local geometry close to 
energy minima   yields significant dynamical information.

For discrete systems, all configurations
within subsets of the state space (pockets) centered around local 
energy minima can be counted\cite{Sibani93a}. 
  In addition, combinatorial techniques have 
  been used to find all low-lying minima
  of small systems\cite{Klotz94a},   and different approaches
have been devised  to coarse grain the configuration
space into a space of more manageable
size\cite{Becker97a,Wales98a,Klotz98a,Schon96a,Heuer97a}.

For clusters and glassy systems with continuous configuration
spaces the main emphasis has been on stochastic methods for
finding  energy minima and  the
lowest saddles which connect them\cite{Berry93a}.
The vibrational modes 
of the systems can then
 be described within a harmonic approximation of
the potential close to the minima. 
 
Low temperature MC/MD-simulations have also been utilized
to  probe the vibrational and
relaxational behaviour on short time scales in the neighborhood
of "frozen-in-networks"\cite{Aihara96a,Olig97a,Olig99a}. 
At higher temperatures, these methods  yield 
insights into the dynamics of the glass transition and aging
phenomena\cite{Lewis91a,Kob97a}. In addition,   
the large scale distribution of 
local minima\cite{Weber85a,Valle92a}
can be  sampled by quenching out of the melt.
 
While knowing the minima and their
connecting saddles  is useful in  predicting
e.g. whether a given system will be glass forming or
not\cite{Ball96a,Wales98a,Heuer97a}, quantitative
knowledge  of the density of states of model
glasses is still rather scanty. 
This information can be
useful in clarifying the connection between structural
glasses and systems with quenched randomness as e.g.
spin-glasses and help to provide a unified picture of
the energy landscapes of complex systems.

With this purpose in mind   we present 
in this Letter large-scale  exhaustive enumeration 
of microscopic states  for on-lattice 
3-D network models of 
glassy   systems \cite{Elliott90a,Gutzow95a,Schon98a}.
 By placing the
system on a lattice, a procedure  successfully employed for 
polymers\cite{Kremer88a,Binder95a}, we can 
analyze in detail regions of
configuration space containing many thousands of local
minima, each minimum representing a topologically
different network.

Our analysis shows that the local trap geometry can
be characterized rather simply. The discussion
 deals with the  connection between 
the geometry and dynamical features of 
glasses, including  thermal metastability,
domain formation and cooling behavior.

{\em \bf   Model.} 
The main structural feature of  covalently
bonded  glasses as e.g. amorphous SiO$_2$  
close to the transition temperature
is  their topology. Accordingly,  such
systems are often modeled as random
networks of building blocks\cite{Zallen83}
where links represent the  covalent bonds.

Since we  are  interested in  configurational changes
which take place on time and length scales
considerably larger than those appropriate
for the vibrational degrees of freedom of
the system, we can put our network on a grid
with a lattice spacing $a$ of the order of 
the bond lenght. The harmonic vibrations cannot be described 
 at this level of  resolution. 
 Furhermore, the energy minima of the discrete
model and the corresponding    
positions of the atoms   will be somewhat 
changed in comparison with 
the   off-lattice network. 
However, overall geomerical features of the
energy landscape  as e.g. the distribution of 
accessible minima and barriers  are not likely to be 
seriously affected.

 We use cubic lattices with lattice constant $a=1$,
periodic boundary
 conditions and a repeated cell of
linear size $S =10$ and volume $V = S^3$.
The number of atoms is $N_A$ and the density  
is $\rho = N_A/V$. 

There is an arbitrariness in the choice of the
elementary moves which define the dynamics on the
energy landscape and hence its connectivity. 
We were guided by the simple physical
consideration that a single move should involve
a change of coordinates which is small.
Each elementary move is therefore taken 
as a shift of the position
of one atom to one of the six adjacent grid  points.
With this definition each state space configuration
has $6 N_A $  neighbors in the landscape.

Our  interaction potential is constructed
such that a well defined crystalline ground state 
is present at a certain density. At the same time
it  allows for many other stable
configurations  with no crystalline structure.
The potential vanishes  except for nearest neighbors,
 which are  atoms closer than $r=3.5$.
 The nearest neighbor interaction
   consists of  a two-body and a three-body term. 
The  radial dependence of the former has a hard-core
for $r<1.6$, and   equals 
$V_2(r) = (r-1.97)(r-3.5)^2(r-7)$, 
for $1.6<r<3.5$.  This potential  
has a minimum at  $r \approx  2.2$.

The three-body term  $V_3(\theta)$ containing 
 the angular dependence is infinite
for angles below $40^o$. In the range
$[40^o,80^o]$ it is strongly
 repulsive and given by
%\begin{equation} 
$V_3(\theta) = -2.9 \times 10^{-7} (\theta - 80)(\theta - 180)^2 \theta$.
%\end{equation}
In  the range $[80^o,180^o]$  the potential
has  a shallow broad minimum close to 
 $120^o$ and vanishes at the 
high end of the interval, as expressed 
by the form: 
$V_3(\theta) = -5.8 \times 10^{-8}(\theta - 80)(\theta - 180)^2 \theta$.
  The optimal local
 coordination in the dense phase is the result
  of the interplay of $V_2(r)$ and $V_3(\theta)$,
the former favoring high coordination (up to
 six neighbors in an octahedral arrangement),
  and the latter one favoring a trigonal
planar environment. 
As a consequence, four-fold (tetrahedral)
 coordination is favored energetically,
  followed closely by five-fold coordination.
Configurations  connected by translations and/or rotations
of the system and hence physically equivalent  are
identified and counted  once.

 {\em \bf Results.}  
The  parallel version of the lid-algorithm 
 utilized for our exhaustive searches
 is  detailed in Ref.~\cite{Sibani98b}. 
The   basic idea of how  the method works
is best  explained by referring to the data 
shown in   Fig.~~1.
For concreteness we shall focus on   the
cone-like structures pertaining to the 
 $50$ atom system.
 The algorithm starts at the reference
configuration represented by the apex of
 the {\em highest lying} cone and visits all
 configurations which can be reached
by a succession of elementary moves never
exceeding an energy barrier $L$. The latter appears as
 the vertical
distance from the apex along the cones' symmetry axis.
 The volume ${\cal V}(L)$
of a pocket of depth $L$ is then  the number of configurations
reachable below energy $L$. In the figure, the width
 of the cone at distance  $L$ from the apex is proportional
to the logarithm of ${\cal V}(L)$.  As we  see, 
the accessible volume is   almost exponentially
 increasing with  the barrier. At some value
$L_{max}$ of the barrier, which in the case of
the highest  pocket  is  
 $\approx 0.5 eV$, the algorithm finds a state of energy
lower than the lowest energy so far. This
state is adopted as a new reference state,
 and the counting starts afresh. By iterating this
 procedure the algorithm  generates a succession of 
cones  located at progressively lower energies.
Each of these is the graphical rendering of a state-space pocket.
We emphasize that, while the counting
within each pocket is exhaustive,
 the path joining  one pocket to
the next (symbolized by the lines connecting the cones)
 is just one amongst the  many possible.
The  largest pockets contain 
 $\approx 10^5 - 10^6$   states. 
A second quantity of importance is the
 local density of states ${ \cal D}(E;L)$,
 describing  the energy distribution
of the states in a pocket of
 depth $L$.  For some
representative instances   
${ \ln\cal D}$ is plotted versus $E$,
where $0\leq E \leq L$, in the same way
as ${\cal V} (L)$ was plotted versus $L$.
This  results in the   conical shapes 
enclosed in the dashed boxes   of Fig.~1,
which  show that    ${ \cal D}(E;L)$ is also close 
to an exponential function of its argument.
A more detailed analysis shows that the 
curves have a slightly concave appearance in 
a semilog plot.

Figure~1 may convey the false impression
that  the pockets are 
devoid of internal structure.
 In reality, 
they all contain a multitude of
local minima, each surrounded by its own basin.
In fact, the number of accessible minima and
 their  density also grow approximately
exponentially as a function of $L$ and $E$
(for fixed $L$), respectively. Part of
 this rich structure is made visible
in Fig.~2, where we consider one
of the previously depicted pockets (deepest
minimum for $N_A = 45$)
 in more detail.
The local density of states of a pocket (top curve) 
and six of its subpockets  are  displayed.   
We remark that   each sub-pocket contains many local minima,
and all the local densities of states have a similar,
close to exponential appearance.
We have also investigated the distribution in energy
of the local minima in each pocket finding shapes similar
to that of the local  density of states, except for  stronger
flattening  close to the lid energy.  
 
In a   harmonic approximation of the vibrational modes
within a local minimum,
one would expect a power-law form of the   density of states.   
The  present  local densities of states, however, pertain
 to  a rather different
situation, i.e.  the much larger
length and time scales of excitations associated with changes in 
bonding topology. To 
check whether a power-law could nonetheless describe the  
data, we shift the abscissa with the energy of the lowest
minimum in each valley and plot the data in a log-log plot.
The  inset of Fig.~2 shows two of the curves  of the main panel,
plotted in the fashion just described.  

Finally, we  considered the radial and angular
 distribution functions $G(r)$ and $P(\theta)$
for networks of different densities.
 In Fig.~3, we depict $P(\theta)$ averaged
over the lowest minima for each density.
The fact that the strain in the network
 increases with density is revealed  by the
increasing predominance of `bad' angles. 

The   acquisition
process for all of our data  took of the
 order of one year of calculation,
running in parallel on $8$ processors
on a  SGI Onyx computer.

{\em \bf Discussion.}  
The local geometrical structure of each pocket can   be described 
in terms of available volume, local density of states, and local density
of minima.
Even though these quantities vary somewhat  from pocket to
 pocket, they  nevertheless  share important features.
  Approximately, the local density of states within a pocket of depth $L$ is 
${\cal D}(E,L) \approx C(L) \exp(E/T_g)$. The parameter $1/T_g$ characterizes the average slope of $\cal D$ in a semilog plot and 
  the function $C(L) $ describes
 the effects of side-valleys joining the main valley as the lid grows.  
 Integrating ${\cal D}$ with respect to $E$ from $E=0$ up to $E=L$ yields 
 ${\cal V}(L) \propto C(L) \exp(L/T_g) $. Since,
  as seen in Fig.~1, ${\cal V}(L)$ also grows almost
   exponentially as a function of $L$, we conclude  
 that  $C(L)$ must itself be an exponential or equal to a constant.
  In the latter case   
  all the states   added by increasing the lid  would  have energies
    very close to the lid itself, 
   which in turn implies that the local minima
are   shallow or lacking altogether.
  In fact, however,   deep sub-pockets frequently  appear
as the lid is increased.  
Since  these side-pockets have
fewer states than the   main pocket, but 
 the same overall shape  with  a lesser
  barrier to a deeper minimum, they are  
  approximately  scaled-down versions of their parent. 
   Referring to Fig.~1 we note that
  this  scaling property also seems to apply
    when moving one level up from the individual
    pockets to the larger scale structure of the
     whole landscape. This is because  
    a high-lying pocket
   is by construction a subset  of a lower-lying one
    and because  all pockets within a certain energy range 
 seem to have similar shapes independently of the
   energy of their lowest minimum.   All in all,
  this is strongly suggestive of a
   self-similar structure in the energy landscape.
Of course, to rigorously check self-similarity one should extend
the  analysis 
to several additional levels of nesting, a task 
 which falls outside the scope of this investigation. 
 From what we can glean so far concerning the largest energy scale, 
 where we also would include very high-lying (possibly physically 
non-relevant) minima,
there appears to be a weak trend to higher values of $T_g$ for
 the high-lying pockets. But since these narrow pockets usually appear
to be relatively small compared to the rest of the system, their influence
on the dynamical properties should not be very strong. 
Quite generally, the information gained from these exhaustive
searches allows the construction or/and validation of 
mesoscopic models of complex system dynamics 
\cite{Sibani89a,Bouchaud94a,Uhlig95a} which rely on assumed
geometrical properties of state space. 
Since  our  pockets are metastable    below
the trapping temperature $T_g$
one can construct and 
 use the connectivity
matrix of small pockets ($< 10^4$ states)
   to obtin direct insight in the low-temperature
 relaxation behavior\cite{Sibani93a}.  
 
For  $T>T_g$, a hypothetical  quasi-equilibrium distribution
within the pocket would be very  skewed in favor of  states
with energy close to the lid\cite{Sibani93a,Sibani98a,Schon97b}. 
Banning unlikely kinetic mechanisms
which would prevent the system from escaping the trap,
 this implies a loss of
metastability at $T \approx T_g$, where the trap 
effectively 
disappears from the
landscape. A purely exponential
density of states gives a singularity in the specific heat when 
the temperature approaches $T_g$ from the low side. (The high 
side is irrelevant since the system has then left the pocket 
and new unchartered parts of state space must enter the calculation). 
Such a peak in $C_V$ is also observed experimentally at the glass transition
temperature, and commonly associated with a configurational entropy
\cite{Elliott90a,Gutzow95a}. The finite size of the pocket and
corrections to the purely exponential behavior smoothen the 
transition of course.
  
The value of $T_g$ decreases with the 
system  {\em  real space  volume} $V$\cite{Schon98a,Sibani98a}. 
In the glass case,  a simple free volume analysis\cite{Schon98a} 
indicates that 
$T_g \propto 1/(V(1-\rho V_A))$, where $V_A$ is the volume/atom
in a dense strain-free network and $\rho$ is the particle density.
The fact that $T_g$ vanishes in the limit of infinite
volume does not imply that trapping does not take place
in large systems. Indeed, a large system will not be able 
to equilibrate internally at low temperatures, 
and the interpretation
of $T_g(V)$ as a trapping temperature is then   no longer
appropriate. Instead one may envisage that a large system
may  form domains of strongly correlated 
atoms, similarly to the behavior observed in
spin-glasses\cite{Fisher88,Kisker96,Andersson96}. The preceding analysis
applied to these domains leads to the conclusion
that the domain size is limited by an entropic
mechanism: for a given temperature $T$
a domain can only  be stable  if $T_g(V_{eff}) \leq  T $.
Domains can grow up to the size at which  
equality holds  in the above relation.  
 
If the system is annealed  from  high temperatures
a  slower cooling rate will arguably lead to 
a larger size  $V_{eff}$ of locally equilibrated domains,
and hence to a lower glass transition temperature, which
 agrees  with the experimental  behavior of glasses
under cooling\cite{Elliott90a,Gutzow95a}.

The basic condition for the above   scenario,
i.e. the approximate exponential growth of local densities of states
in locally ergodic regions of the energy landscape,
is fulfilled in a number of additional systems 
that show complex dynamics and glass-like behavior: the TSP\cite{Sibani93a},
spin glasses\cite{Sibani98a}, and lattice polymers\cite{Schon97a}.
Thus, we suggest that  a loss of metastability in spatially
localized traps with exponential densities of states, 
i.e.  mainly  an entropic mechanism, is an important factor 
in the glass transition and in  cooling
rate dependent dynamic effects.

\noindent{\bf Acknowledgments}\\                
Funding was kindly provided by the DFG via SFB408 and by a block grant 
from  Statens Naturvidenskabelige Forskningsr\aa d.

\newpage

 \begin{figure}[t]
 \vspace{-0.5cm}
% \normalfigure{Fig1}
  \centerline{\psfig{figure=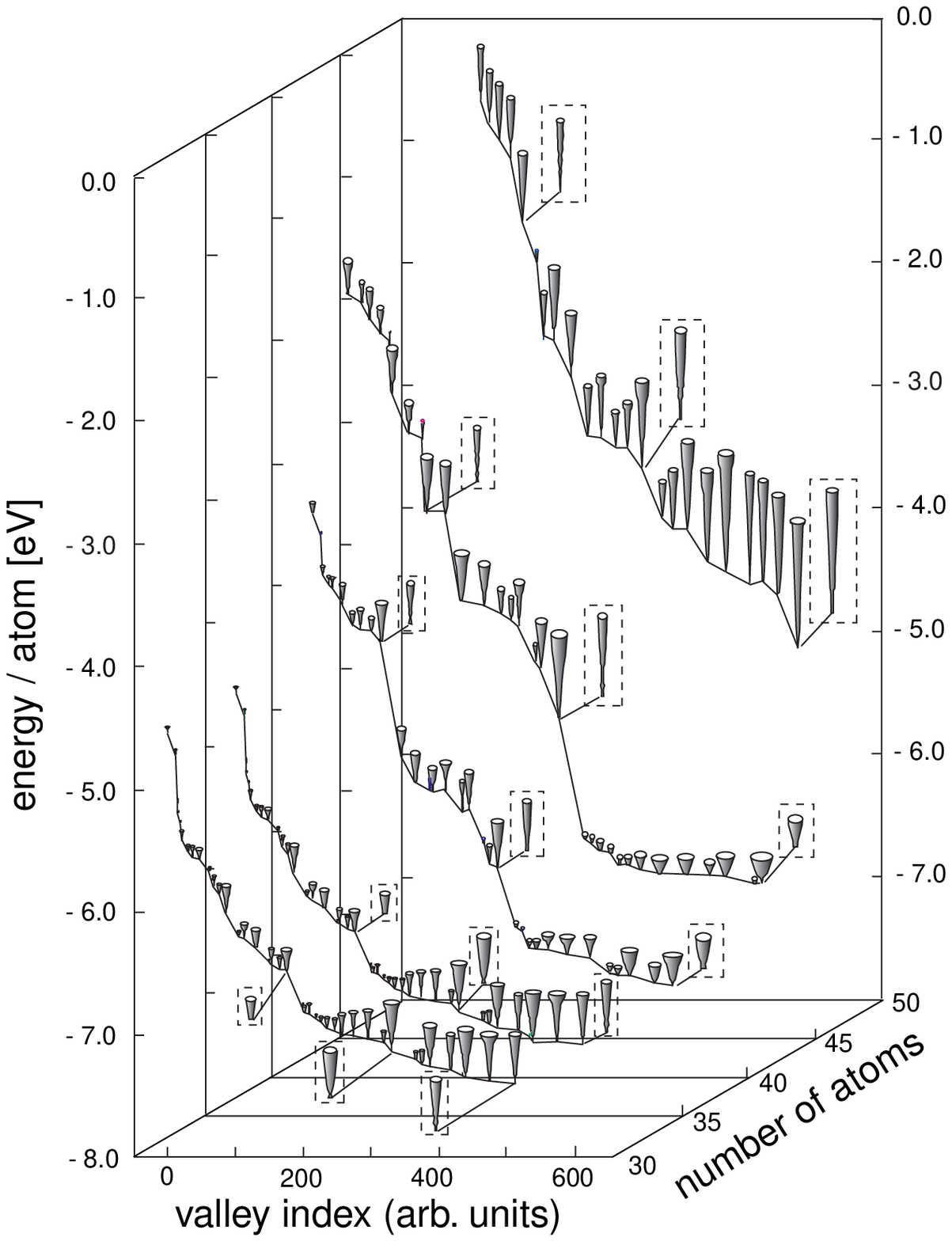}}
 %\vspace{0.1cm} 
 \end{figure}
\begin{figure}
\caption{  
A birds-eye  view of the energy landscape of the glass model, 
where pockets appear  as  cone-like structures. The width 
of the cones at energy $L$ relative to the position of the
apex   is proportional to the logarithm 
of the  locally accessible state-space volume ${\cal V}(L)$.
The insets use the same graphical 
rendering  but with the cone-width at energy $E$
proportional to the logarithm of    $  {\cal D}(E;L_{max}) $,
 the local density of states available in 
the selected pockets.  
}
\end{figure}

\newpage
 
 \begin{figure}[t]
% \normalfigure{Fig2}
 \centerline{\psfig{figure=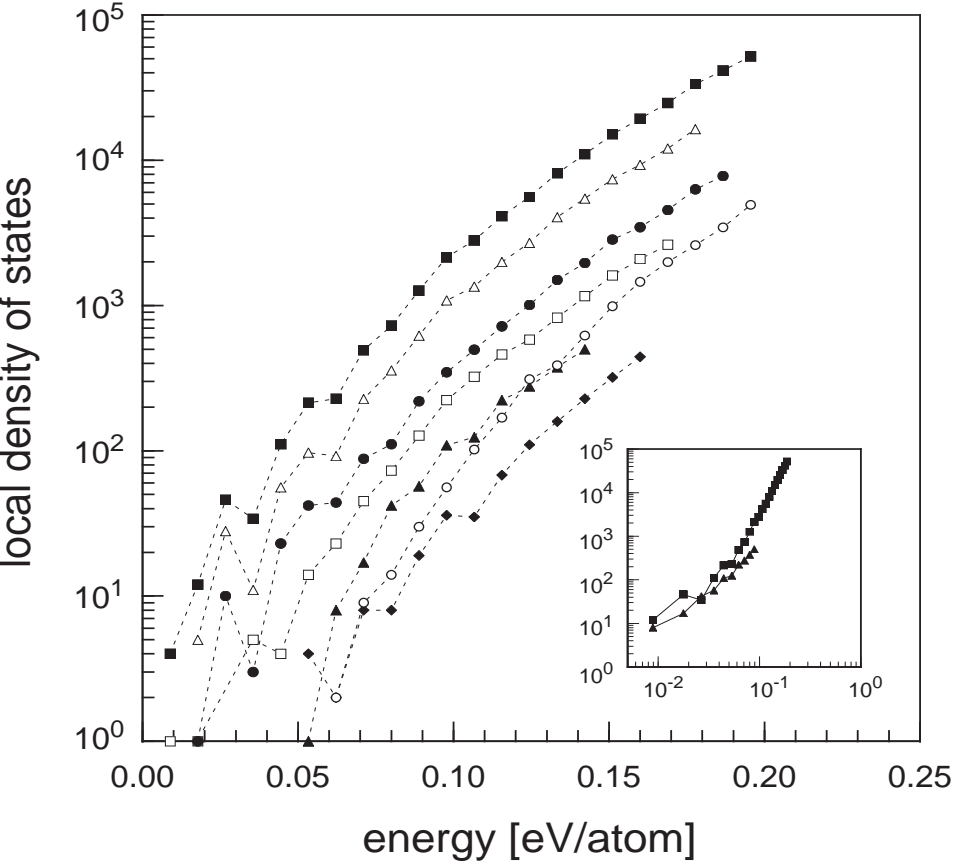,height=18cm,width=14cm}}
 \vspace{-1.5cm}
\caption{ 
The local density of states of several  nested sub-pockets
 of a system with $45$ atoms. The top curve (black squares)
describes the local density  of all
states in the full pocket up to an energy of $\approx 0.22$ eV$/$atom.
Each other curve describes the local density of states of a sub-pocket
merging into the main valley approximately at the energy where the
curve ends, within the accuraqcy given by the binning of the energy
axis.
E. g., the second  curve (empty triangles) corresponds to 
 a subpocket merging at 
$\approx  0.19$ eV$/$atom.
 Note that the growth of the  local densities of states of the sub-pockets closely parallels the one of the main pocket.
The  inset shows  the same data as the top  
and bottom curves of the
main panel, plotted on  log-log scales and
with the  abscissa values shifted, such that the
lowest energy in the appropriate (sub)valley  
is located at the origin.   
}
\end{figure}

 \newpage 

\begin{figure}[t]
%\normalfigure{Fig3}
  \centerline{\psfig{figure=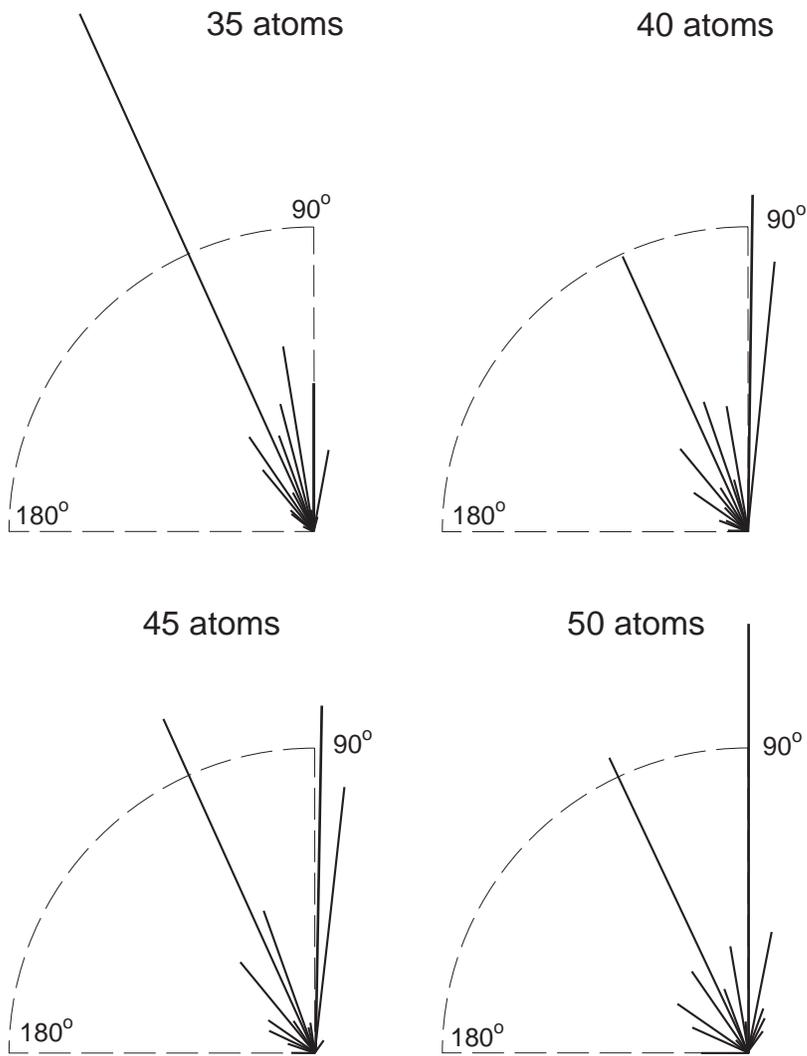}}
 \vspace{2cm}
\caption{ 
The length of each ray is proportional to the frequency of the
corresponding bond angle in the configurations of the  deepest energy
minima of the three lowest pockets for systems of $35, 40, 45, 50$  atoms,
respectively. The $30$  atom system (not shown) is similar to the $35$ atom one,
exhibiting even more (ideal) tetrahedral surroundings.
The preferred bond angle is close to $110^o$, due to the interplay
of the two-body- and the three-body-terms in the interaction potential.
Note the increasing predominance of energetically unfavorable angles as the
atom density increases.
} 
\end{figure}

\end{document}